\documentclass{ws-ijmpe}
\usepackage{epsfig}
\usepackage{graphicx}
\def\aA{$\alpha$-nucleus }
\def\AA{nucleus-nucleus }

\def\ac{$\alpha+^{12}$C }
\def\ext{$\alpha$+$^{12}$C$^*$\ }

\begin{document}

\markboth{D.T. Khoa}{Probing the isoscalar excitations of $^{12}$C with
inelastic alpha scattering}

\catchline{}{}{}{}{}

\title{PROBING THE ISOSCALAR EXCITATIONS OF $^{12}$C \\
WITH INELASTIC ALPHA SCATTERING}

\author{DAO T. KHOA\footnote{Email: khoa@vaec.gov.vn}}
\address{Institute for Nuclear Science and Technique, VAEC \\
 179 Hoang Quoc Viet, Nghia Do, Hanoi, Vietnam}

\maketitle

\begin{history}
\received{(received date)}
\revised{(revised date)}
\end{history}

\begin{abstract}
The robust (spin and isospin zero) $\alpha$-particle remains one of the best
projectiles to probe the nuclear isoscalar excitations. In the present work, a
microscopic folding model analysis of the \ac inelastic scattering to the 2$^+$
(4.44 MeV), 0$^+$ (7.65 MeV), 3$^-$ (9.64 MeV), 0$^+$ (10.3 MeV) and 1$^-$
(10.84 MeV) states in $^{12}$C has been performed using the 3-$\alpha$
resonating group method wave functions. The isoscalar transition strengths of
these states were carefully studied based on the coupled-channel analysis using
the microscopic folded form factors. A correlation between the weak binding
and/or short lifetime of the excited state and absorption in the exit channel of
inelastic scattering has been established.
\end{abstract}

\section{Introduction}
Given the well established $\alpha$-cluster structures of some excited states of
$^{12}$C, the low-lying isoscalar (IS) excitations of $^{12}$C become a subject
of significant interest recently \cite{Fre07}. In particular, the isoscalar
0$^+_2$ state at 7.65 MeV in $^{12}$C (known as the Hoyle state) has been
studied extensively due to its vital role in the stellar synthesis of Carbon.
Although this state was identified long ago in the inelastic \ac scattering
\cite{Spe71,Smi73,John03} and inelastic electron scattering \cite{Stre70} as an
isoscalar $E0$ excitation, our knowledge about its unique structure is still far
from complete \cite{Fre08}. Since the Hoyle state lies slightly above the
$\alpha$-decay threshold of 7.27 MeV, its wave function turns out to have a
dominant three $\alpha$ clusters component, which has been confirmed, e.g., by
the Resonating Group Method (RGM) calculations \cite{Ueg77,Kam81}. Quite
interesting is the condensate structure suggested in Refs.~\cite{Toh01,Fun03}
where the three $\alpha$ clusters were shown to condense into the lowest
$s$-state of their potential and form, therefore, a Bose-Einstein condensate
(BEC). Such a BEC structure of the Hoyle state was shown \cite{Che07} to be
mixed also with the molecular $^8$Be$+\alpha$ configuration.

In general, to validate conclusion made in the structure calculation, the wave
functions must be carefully tested in the study of nuclear reactions. Since the
spin- and isospin zero $\alpha$-particle is a very good projectile to excite the
nuclear IS states, the 3-$\alpha$ RGM wave function of the Hoyle state obtained
by Kamimura \cite{Kam81} has been recently used in a folding model analysis
\cite{Kho08} of the inelastic \ac scattering and realistic $E0$ transition was
deduced. In the present work we extend this approach to study also other IS
excitations of $^{12}$C like 2$^+$ (4.44 MeV), 3$^-$ (9.64 MeV), 0$^+$ (10.3
MeV) and 1$^-$ (10.84 MeV) states, using the RGM wave functions \cite{Kam81}.

\section{Double-folding calculation and coupled-channel scheme}
In general, the \aA form factor (FF) contains all structure information of the
nuclear state and it is, therefore, desirable to have FF evaluated in the
microscopic folding model using an appropriate effective nucleon-nucleon (NN)
interaction and realistic wave functions for the $\alpha$-particle and target
nucleus, respectively. The main ingredients of the folding approach and
coupled-channel (CC) formalism for elastic and inelastic \AA scattering can be
found, e.g., in Refs.~\cite{Kho00,Sat83}.

In this work, we have used the \emph{complex} density dependent CDJLM
interaction constructed \cite{Kho08} based on the Brueckner Hartree-Fock results
for the nucleon optical potential (OP) in symmetric nuclear matter. The nuclear
wave functions obtained in the RGM method by Kamimura \cite{Kam81} have been
used to generate nuclear densities for the folding calculation of the complex OP
and transition FF for the \ac system. These RGM wave functions were proven to
give consistently a realistic description of the compact structure of the ground
state $0^+_1$ and the first $2^+$ and $3^-$ states, as well as the dilute
structure of the $0^+_2$ (Hoyle state) and 1$^-$ state (see Table~1).
\begin{table}[ht] \tbl{Characteristics of the IS states of $^{12}$C
 under present study. Calculated values are those taken from the RGM
 calculations \protect\cite{Kam81}. $S_\rho$ is the scaling factor
 of the complex inelastic FF used in the renormalized DWBA
 calculation (see Fig.~2). $S_{\rm W}$ is the ratio of the absorption strength
 in the exit channel to that in the entry channel of inelastic \ac scattering.}
{\begin{tabular}{ccccccccc}\toprule
 $J^\pi$ & $E_{\rm cal}$ & $E_{\rm exp}$ & $B(EJ)_{\rm cal}\ ^\star$ &
 $B(EJ)_{\rm exp}\ ^\star$ & $\tau_{\rm exp}$ & RMS$_{\rm cal}$ & $S_\rho$ &
 $S_{\rm W}$ \\ &
(MeV) & (MeV) & ($e^2$fm$^{2J}$) & ($e^2$fm$^{2J}$) & (sec) & (fm) & &
 \\ \colrule
 2$^+$ & 2.77 & 4.44 & 46.5 & $40\pm 4$ & $6\times 10^{-14}$ & 2.38 & 1.00
 & 1.00 \\
 0$^+$ & 7.74 & 7.65 & 6.62 & $5.4\pm 0.2$ & $8\times 10^{-17}$ & 3.47 & 0.55
 & 3.27 \\
 3$^-$ & 8.14 & 9.64 & 872 & $610\pm 90$ & $2\times 10^{-20}$ & 2.77 & 0.75
 & 1.75 \\
 0$^+$ & 14.0 & 10.3 & 6.33 & -- & $2\times 10^{-22}$ & 3.14 & 0.45 & 4.85 \\
 1$^-$ & 10.8 & 10.8 & 4.16 & -- & $2\times 10^{-21}$ & 3.39 & 0.20 & 8.60 \\
\botrule
\end{tabular}}
\begin{tabfootnote}
$^\star\ M(EJ)$ in $e$~fm$^{J+2}$ for 0$^+$ and 1$^-$ states.
\end{tabfootnote}\label{t1}
\end{table}
For the Hoyle state, the RGM wave function has been shown \cite{Fun03,Fun06} to
be very close to the BEC wave function. To take into account coupling between
different scattering channels, the RGM densities were constructed for both the
diagonal and nondiagonal transitions between states under study, including the
$E2$ reorientation of the $2^+$, $3^-$ and 1$^-$ states. In total, 23 diagonal
and nondiagonal densities were used to calculate the \aA OP and inelastic FF for
the CC equations, with the coupling scheme shown in Fig.~1.
\begin{figure}[ht]
 \begin{center}
  \vspace{-3cm}\hspace{0cm}
 \includegraphics[angle=0,scale=0.40]{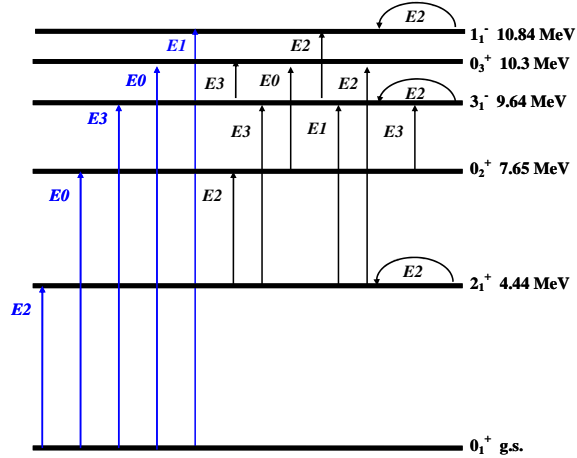}
  \vspace{-1.50cm}
\caption{Coupling scheme used in the CC equations for elastic and inelastic \ac
scattering.}
   \end{center}\label{f1}
\end{figure}
One can see from Fig.~1 that the two-step excitation of the IS states are
treated in equal footing with the direct excitation. In particular, the two-step
excitation ($0^+_1\to 2^+\to 0^+_2$) of the Hoyle state is the exact inverse of
transitions taking place in the stellar Carbon production. Since the $E\lambda$
transition strengths for some states are well determined from experiment, we
have slightly scaled the RGM densities to match the calculated $B(E\lambda)$
values with the data for $B(E2,0^+_1\to 2^+),\ B(E2,2^+\to 0^+_2)$ and
$B(E3,0^+_1\to 3^-)$, so that the corresponding inelastic FF can be reliably
used in the CC analysis of the inelastic \ac scattering. For the $0^+_2$ and
$1^-$ states, given a good description \cite{Kam81} of the $(e,e')$ data by the
corresponding RGM transition densities, we kept them unchanged in the folding
calculation. All the CC calculations have been performed using the code ECIS97
written by Raynal \cite{Raynal}.

\section{Results and discussion}
Although many experiments on inelastic \ac scattering were done (see, e.g.,
Refs.~\cite{Spe71,Smi73}) in the last century, inelastic scattering to the
$0^+_3$ (10.3 MeV) and 1$^-$ (10.8 MeV) states have been measured with high
precision only recently at the $\alpha$ energy of 240 MeV \cite{John03}. We
concentrate, therefore, on the 240 MeV data in the present work. To fine tune
the strength of complex CDJLM interaction, the real and imaginary elastic folded
potentials were slightly renormalized by an optimal fit to the elastic
scattering data. The same renormalization factors were then used to scale the
real and imaginary inelastic folded FF for the inelastic scattering study
\begin{figure}[ht]
 \begin{center}
  \vspace{2cm}\hspace{0cm}
 \includegraphics[angle=0,scale=0.40]{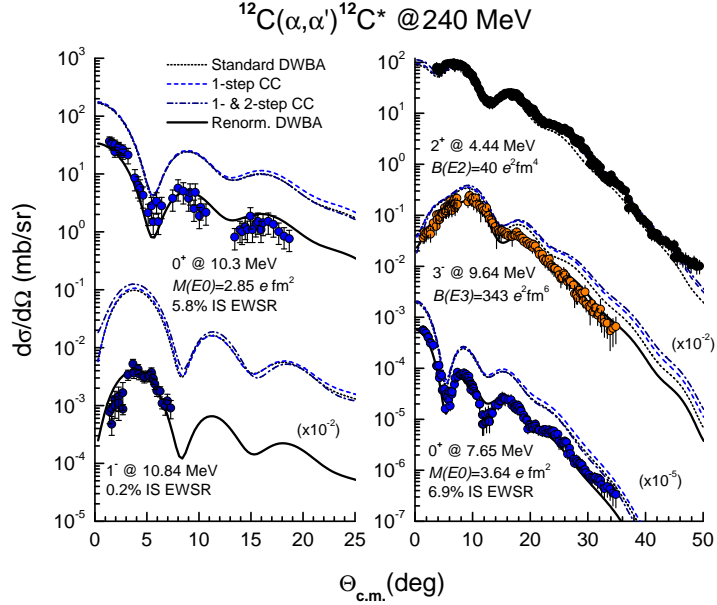}
  \vspace{-2.5cm}
\caption{Inelastic \ac scattering data measured at $E_{\rm lab}=240$ MeV
\protect\cite{John03} for the IS states of $^{12}$C under study in comparison
with the DWBA and CC results obtained with the folded OP and inelastic FF. The
transition momenta (or probabilities) and fractions of the EWSR are deduced from
the renormalized DWBA calculation.}
   \end{center}\label{f2}
\end{figure}
either in the distorted wave Born approximation (DWBA) or in one- and two-step
CC schemes. The DWBA and CC results are compared with the data in Fig.~2. Except
for the $2^+$ state, the calculated cross sections for other states
significantly overestimate the data, especially for the $0^+_3$ and 1$^-$
states. The CC effect by one- and two-step coupling could not help to solve this
problem. To fit the data in the conventional DWBA or CC methods, the complex
inelastic FF needs to be scaled by a factor $S_\rho$ (see Table~1 and the
renormalized DWBA results plotted in Fig.~2). Since the RGM transition densities
were used to reproduce quite well the experimental $(e,e')$ form factors of
these states \cite{Kam81}, such a scaling leads naturally to a ``suppression" of
the IS transition strength, in terms of $B(E\lambda)$ probability or fraction of
IS energy weighted sum rule (EWSR) shown in Fig.~2. It can be seen from Table~1
that the IS strengths deduced from the renormalized DWBA results strongly
disagree with the experimental values. For example, the renormalized DWBA
implies about 6.9\% of the IS monopole EWSR for the Hoyle state, while this
fraction should be around 15\% as deduced from the $(e,e')$ data \cite{Stre70}.
Since the $(e,e')$ data measured by Strehl \cite{Stre70} covers only low
momentum transfers, we compare in Fig.~3 the $(e,e')$ form factors for the Hoyle
state given by the RGM transition density with a set of high precision $(e,e')$
data taken up to high momentum transfers \cite{Cra05,Len06}.
\begin{figure}[ht]
 \begin{center}
  \vspace{1.5cm}\hspace{0cm}
 \includegraphics[angle=0,scale=0.33]{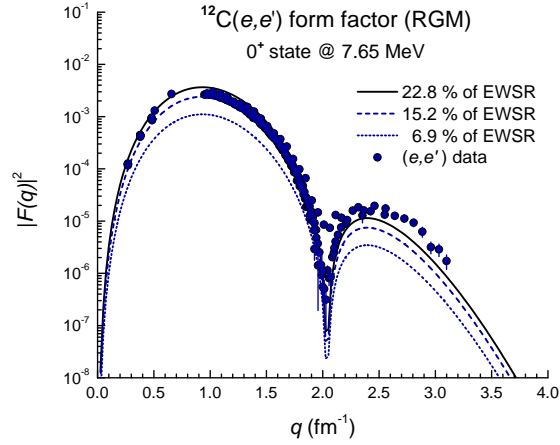}
  \vspace{-2.5cm}
\caption{Inelastic electron scattering FF for the Hoyle state given by the
original RGM transition density \protect\cite{Kam81} (which exhausts 22.8\% of
the IS monopole EWSR) and that scaled to exhaust 15.2\% (as deduced from
$(e,e')$ data at low momentum transfers \protect\cite{Stre70}) and 6.9\% (as
deduced from the renormalized DWBA) of the monopole EWSR. The data are taken
from Refs.~\protect\cite{Cra05,Len06}.}
   \end{center}\label{f3}
\end{figure}
\begin{figure}[ht]
 \begin{center}
  \vspace{2cm}\hspace{0cm}
 \includegraphics[angle=0,scale=0.4]{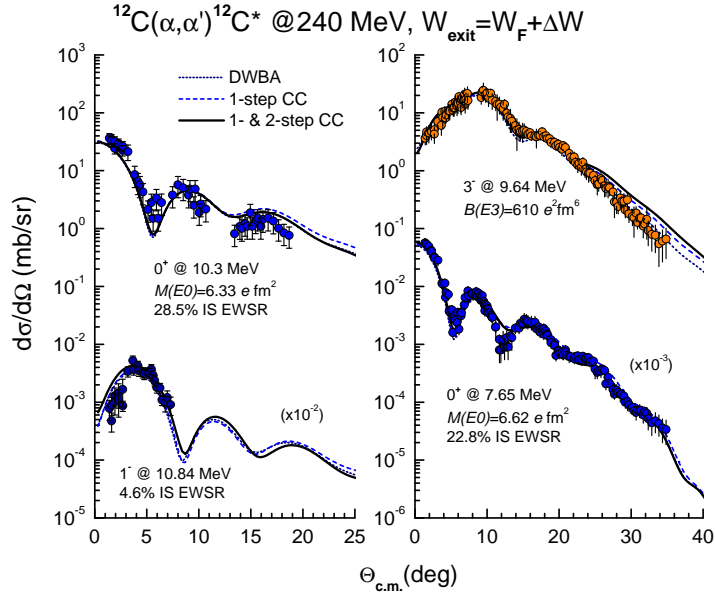}
  \vspace{-2.5cm}
\caption{The same as Fig.~2 but with the enhanced absorption in the exit \ext
channels. The transition momenta (or probabilities) and fractions of the EWSR
are given by the RGM transition densities used in the folding calculation. See
more details in text.}
   \end{center}\label{f4}
\end{figure}
It is easy to deduce from Fig.~3 that the $E0$ transition strength of the Hoyle
state exhausts 15 - 23 \% of the IS monopole EWSR, more than twice the sum rule
fraction found from the conventional DWBA analysis of inelastic \ac scattering
\cite{John03,Kho08}. We have found \cite{Kho08} that an enhanced absorption in
the exit channel (due to the short lifetime and weakly bound structure of the
Hoyle state) leads directly to a suppression the monopole strength in the
inelastic \ac scattering. To explore similar effect for other IS states of
$^{12}$C, we have added to the microscopic folded imaginary OP for each exit
\ext channel a surface term $\Delta W$ with parameters determined from the best
DWBA and CC fit to the inelastic data. In this way, consistent descriptions of
the inelastic \ac scattering and $(e,e')$ data could be reached using the RGM
nuclear transition densities (see Fig.~4). Excepting for the $2^+$ state, the
imaginary OP in the exit \ext channel turned out, as expected, to be much
stronger than that in the entrance $\alpha+^{12}$C$_{\rm g.s.}$ channel (see the
ratio $S_{\rm W}$ of the volume integral of the imaginary OP in the exit channel
to that in the entrance channel in Table~1). By comparing the nuclear RMS radii,
mean lifetimes $\tau$ and ratios $S_{\rm W}$ for the IS states of $^{12}$C under
study, we conclude that there exists a correlation between the diluteness,
lifetime (or weak binding) of the excited state and absorption in the exit
channel. Since most of the candidates for $\alpha$-cluster states are weakly
bound, this absorption effect must be taken into account in the future studies
of cluster states excited in the inelastic \aA or \AA scattering.

\section*{Acknowledgements}

The author thanks Prof. W. von Oertzen and Dr. T. Neff for helpful
communications and Prof. M. Kamimura for the revised parametrization of the RGM
densities from Ref.~\cite{Kam81}. This work has been supported, in part, by the
Alexander von Humboldt Stiftung of Germany and Natural Science Council of
Vietnam.

\end{document}